\documentclass[a4paper,12pt]{article}
\usepackage[utf8]{inputenc}
\usepackage{amsfonts,amsthm,amsmath,amssymb,graphicx,hyperref,youngtab,color}
\numberwithin{equation}{section}
\usepackage{cite}

\newcommand{\be}{\begin{equation}}
\newcommand{\ee}{\end{equation}}
\newcommand{\bea}{\begin{eqnarray}}
\newcommand{\eea}{\end{eqnarray}}

\newcommand{\ba}{\begin{eqnarray}}
\newcommand{\ea}{\end{eqnarray}}

\newcommand{\la}{\langle}
\newcommand{\lb}{\rangle}

 \def\la{\langle}
 \def\lb{\rangle}

\title{The Non-melonic Sector of Tensor Models and Gravity}
\author{Pablo Diaz\thanks{pablodiazbe@gmail.com}\\
{\small \emph{EUPT, Departamento de Matemática Aplicada, Universidad de Zaragoza.}}\\
}

\begin{document}

\maketitle
\begin{abstract}
The melonic sector has been proven to be dominant in tensor models at large N. This is true as long as the observables we consider, composites of $2n$ tensors, are small. That is, if $n\ll N$. In this paper, I argue that, in order to recover geometries (and then gravity) in the continuum limit, $n$ must grow like $N$. In that case, I provide examples where non-melonic contributions overcome the total sum in the computation of the expectation value of certain observables.
\\

\end{abstract}
\newpage
\tableofcontents

\section{Introduction}
Understanding the quantum nature of spacetime is one of the most fascinating and intriguing challenges of theoretical physics nowadays. As opposed to other branches of physics, like quantum mechanics, which where developed in agreement with experiments, the extreme high energies we would need to test quantum gravity effects (of the order of the Plank energy) make it impossible to check or rule out ideas based on lab facts, and we can only rely on the mathematical consistency of the proposals. The most basic check any theory of quantum gravity must pass is to prove that Einstein gravity is recovered at the classical limit.

Although there is not a unified framework where to study quantum gravity, some approaches have brought insights into it.  From different perspectives and with some overlaps we can approach quantum gravity from string theory, holography, non-commutative geometry, canonical quantum gravity, spacetime triangulations, tensor theories... It is likely that one of them (or a new one) prevails in the future. Probably because it will offer a clearer picture or because calculations are more attainable. I do not find, however, any philosophical conflict in the fact that there exist different valid options. At the end of the day, any model can give an accurate description  of a physical phenomenon as long as it captures the relevant degrees of freedom, and mathematics is rich enough to accommodate them into different frameworks.

One of the exciting features of tensor theories is the idea of ``spacetime emergence''. This is not the case of string theory, for instance, for which a background where the strings live is assumed {\it a priori}. Historically, the idea of describing $d$-dimensional spacetime by means of tensor models with $d$ indices, comes from the unquestionable success of matrix theories to describe 2-dimensional gravity \cite{earlierTM} in the limit where the size of the matrices, $N$, goes to infinity. However, the tensor models that were proposed at the beginning \cite{random1,random2,random3} were pathological at large $N$, and soon the efforts in this direction declined. It was not until 2009, with the arrival of color tensor models \cite{color1,color2}, and its well-defined $1/N$ expansion \cite{1N1,1N2,1N3}, that the subject was recovered \cite{GR,Diaz:2017kub,Diaz:2018xzt,Diaz:2018zbg,Diaz:2018eik,Itoyama:2019sos,Kemp:2019log,BenGeloun:2017vwn,deMelloKoch:2017bvv,Itoyama:2017wjb,Itoyama:2018but,Delporte:2018iyf} and tensor models became solid candidates for the description of quantum gravity effects. For a more comprehensive bibliographic information see \cite{Klebanov:2018fzb}, and the references therein.

In the $1/N$ expansion of color tensor models it was soon realized that a sector of the theory, the so-called melonic sector since the Feynman diagrams look like melons, dominate the partition function and, as a consequence, it also dominates the expectation value of the observables \cite{Bonzom:2011zz}.  This is why most of the studies in tensor models have focused on the melonic sector so far. The melonic sector of tensor models is analogous to the planar sector in matrix models. There are exciting features about melons. For instance, their simplicity and their pattern of proliferation permits exact perturbative calculations \cite{Bonzom:2011zz}. As a drawback, however, the number of melon observables seems to count branched polymers \cite{Bonzom:2011zz,Bonzom:2012hw}  (tree diagrams) and not manifolds, which kind of jeopardizes the one-to-one association of tensor observables and geometries, as expected at large $N$.   

  The connection of tensor models and geometry is made as follows. Observables in tensor models are composites of $n$ copies of a tensor $T$ contracted with $n$ copies of its complex conjugate $\overline{T}$ in a specific manner. Tensor observables are in one-to-one correspondence with piece-wise linear (PL) manifolds since any pattern of contraction can be interpreted as dictating how to glue simplices by their faces to make a PL manifold. So tensor observables are the skeletons of PL manifolds. Remarkably, it was proven that, providing suitable identifications, the dynamics of tensor models matches the dynamics of triangulations ruled by the Regge calculus \cite{Bonzom:2011zz}, whose continuum limit is Einstein gravity. It is a common belief that the continuum limit is achieved when $n\to \infty$, that is, with an infinite number of simplices. This means that only large tensor observables will become geometries.

In this work I question the dominance of the melonic sector in tensor models in the limit that they make contact with gravity.
The key idea is that the limits $N,n\to \infty$ are not to be taken uncorrelated. To support this claim I go deeper in the investigation of the physical meaning of $N$ in tensor models. It is obvious that $N$ is a measure of the degrees of freedom of the model: the greater  the $N$ the more degrees of freedom. However, a method, like diagonalization in matrix theories, to know the true degrees of freedom of a given tensor is not available. Instead, I use the canonical decomposition \eqref{tensordecomp}. The actual degrees of freedom of any tensor seem to get arranged into $d$-uplas that can be naturally interpreted as $d$-vectors living in the tangent space of the limiting manifold. The fact that for manifolds each point carries its tangent space leads me to claim that $n$ must grow like $N$. As a consequence, subleading non-melonic contributions in the computation of the expectation value of an observable, for being so numerous, can overcome the total sum and should not be neglected. 

The paper is organized as follows. Section \ref{MF} contains all the mathematical tools I will need in the discussion.  Specifically, subsection \ref{TM} offers a brief introduction to tensor models and sets the notation; in subsection \ref{R1} the canonical decomposition of a tensor is explained; subsection \ref{GDI} reminds the key concept of G-degree in the $1/N$ expansion and brings a novel formula to calculate them, equation  \eqref{gdegree2}; in subsection \ref{EVGD}, we recall how the expectation value of observables can be computed in terms of the G-degree of graphs. The main discussion of the paper is placed in section \ref{DL}, where the claim that $N$ grows like $n$ is made. As a consequence of it, I show in section \ref{DoM} a couple of examples where the non-melonic contributions are not subleading.  \\

While preparing this paper, an article with similar conclusions but completely different approach appeared  \cite{deMelloKoch:2019lsx}. Using holographic arguments they claim that the ``melonic limit'' does not provide enough structure as to create a new holographic dimension. I believe that the reasons they have found have to do with my arguments in the present paper although a deeper understanding on both sides will be necessary to establish a precise link. In any case, this ``coincidence'' supports the spirit that different approaches can cooperate in the understanding of quantum gravity.

\section{Mathematical framework}\label{MF}
In this section I provide a brief collection of the mathematical tools and notation I will use in the main discussions. 
\subsection{Tensor models}\label{TM}
Colored tensors are tensors with no further symmetry assumed. A $d$-covariant color tensor can be written as 
\begin{equation}\label{dtensor}
 \Phi=\Phi_{i_1i_2\dots i_d}~e^{i_1}\otimes e^{i_2}\otimes \cdots \otimes e^{i_d},
\end{equation}
 where $\{e^{i_k}\}$ form a basis of $\mathbb{C}^{N_k}$, so $i_k=1,\dots, N_k$. This is the most general object. \\
 In this article we are taking all the vector subspaces with the same size, namely $N$. Then, the objects  $ \Phi_{i_1i_2\dots i_d}$ transform under the action of the gauge group $U(N)^{\otimes d}$  as
\begin{equation}\label{unitaryaction}
 \Phi_{j_1j_2\dots j_d}=\sum_{i_1,\dots, i_d}U_1(N)_{j_1}^{i_1}U_2(N)_{j_2}^{i_2} \cdots U_d(N)_{j_d}^{i_d}  \Phi_{i_1i_2\dots i_d},
\end{equation}
where the subscript of the groups have been added to note that they are different copies of $U(N)$.\\
The complex conjugate is a contravariant tensor that transforms as
\begin{equation}\label{unitaryactionconjugate}
 \overline{\Phi}^{j_1j_2\dots j_d}=\sum_{i_1,\dots, i_d}\overline{U}_1(N)^{j_1}_{i_1}\overline{U}_2(N)^{j_2}_{i_2} \cdots \overline{U}_d(N)^{j_d}_{i_d}  \overline{\Phi}^{i_1i_2\dots i_d}.
\end{equation}
Observables in tensor models are invariants under the gauge group\footnote{Note that all through this paper I will talk about ``invariants'', ``observables'' and later about (Feynman) ``graphs'' indistinctively, since there is a one-to-one correspondence among them.}. They are built by contracting the indices of $T$ and $\overline{T}$ of the same color. Using one pair there is only one possible contraction
\be
T_{i_1\dots i_d}\overline{T}^{i_1\dots i_d}.
\ee
If more tensors are involved each pattern of contraction leads to a different invariant. A convenient way of parametrizing the space of invariants built with $n$ pairs is by collections of $d$ permutations of $n$ elements, which dictates the patterns of contraction. The notation of the indices in the general case can get tedious. Without loss of generality I will show a generic invariant built on $n$ pairs of an order-3 tensor $T_{ijk}$, where I have chosen different letters for the three colors to simplify the notation. Then,
\be
\mathcal{O}_{\alpha,\beta,\gamma}=T_{i_1j_1k_1}\cdots T_{i_nj_nk_n}\overline{T}^{i_{\alpha(1)}j_{\beta(1)}k_{\gamma(1)}}\cdots \overline{T}^{i_{\alpha(n)}j_{\beta(n)}k_{\gamma(n)}},
\ee
and the triplet $(\alpha,\beta,\gamma)\in S_3^{\times 3}$ specifies the contraction pattern. As noticed in \cite{GR}, shuffling the $n$ copies of $T$ or the $n$ copies of $\overline{T}$ does not affect the invariant. Thus, invariants are equivalence classes of permutation triplets with the diagonal action of $S_n$ on the right and on the left. In other words, 
\be
\mathcal{O}_{\alpha,\beta,\gamma}\sim \mathcal{O}_{\sigma\alpha\tau,\sigma\beta\tau,\sigma\gamma\tau}, \quad \sigma,\tau\in S_n.
\ee
Invariants for order-$d$ tensors are specified by $d$ permutations and the same equivalence due to shuffling slots holds.  I will refer to them as
\be
\mathcal{O}_{\vec{\alpha}},
\ee
where I have called $\vec{\alpha}=(\alpha_1,\dots,\alpha_d)\in S_n^{\times d}$ to streamline the notation.

The Gaussian expectation value of an observable $\mathcal{O}_{\vec{\alpha}}$ is given by
\be
\la \mathcal{O}_{\vec{\alpha}} \lb=\int d\bar{T} d T \,\mathcal{O}_{\vec{\alpha}}\exp{(-N^{d-1}\,T\cdot \bar{T})}.
\ee
The Wick contractions read
\be
\la T_{i_1 \dots i_d}\,\bar{T}^{j_1 \dots j_d}\lb=\frac{1}{N^{d-1}}\delta_{i_1}^{j_1}\cdots \delta_{i_d}^{j_d},
\ee
and so
\be\label{expectation}
\la \mathcal{O}_{\vec{\alpha}} \lb=\frac{1}{N^{d-1}}\sum_{\sigma\in S_n}N^{C(\sigma \alpha_1)+\dots +C(\sigma \alpha_d)}.
\ee

\subsection{Canonical decomposition of a tensor}\label{R1}
A complex tensor of {\it order} $d$ and {\it size} $N$ is an array of $N^d$ complex numbers. Although the tensor has $N^d$ entries, the number of degrees of freedom might be smaller. Given a tensor, it is interesting to know how many degrees of freedom it has and how they can get organized. In the case of matrices, their eigenvalues are the actual degrees of freedom, and encode the physics. One may ask if we can do something similar in tensors, like diagonalizing them by means of unitary transformations. That is, transforming the tensor in a way that the only non-zero entries occur at $T_{iii}$. The answer is no:  A generic tensor is not diagonalizable \cite{CS}. So diagonalization does not seem to be the question one should be asking tensors in order to count their degrees of freedom. Instead, this job is naturally carried out by the so-called canonical decomposition of the tensor. 

Given a tensor $T$ there is always a possible decomposition
\be\label{tensordecomp}
T=\sum_{r=1}^{R(T)}a^{(1)}_r\otimes a^{(2)}_r\otimes\cdots\otimes a^{(d)}_r,
\ee
where each $a^{(k)}_r$ is a complex $N$-vector. This decomposition is not unique, especially if we do not restrict the value of $R(T)$. There is, however, a minimum value of $R(T)$ for the decomposition. This minimum value is called the {\it rank} of $T$.
The number of degrees of freedom of rank-1 tensors is always 
\be\label{dof}
\# \,\text{dof } (T)=dN-d+1,
\ee
as can be easily computed. Basically, it is the number of degrees of freedom of $d$ complex $N$-vectors except for rescale ambiguities, say,
\be
T=\vec{a}\otimes \vec{b}\otimes \vec{c}=\frac{\vec{a}}{a_1}\otimes \frac{\vec{b}}{b_1}\otimes \vec{c} \,(a_1b_1).
\ee
For arbitrary rank, the decomposition  \eqref{tensordecomp} for minimal $R(T)$, the {\it canonical decomposition} \cite{CC,Harshman}, is found to be unique as long as
\be\label{doftotal}
R(T)\big[d N-d+1\big]\leq N^d,
\ee
as can be checked by counting the degrees of freedom. The quantity on the LHS of \eqref{doftotal} is the total number of (complex) degrees of freedom of the tensor.

Note that the definition of rank I have just described extends to matrices and coincides with the usual concept: rank of the matrix $A$ is the number of its non-zero eigenvalues. However, as opposed to matrices, where their rank is always equal or less than their size, in tensors the rank can be (and typically is) greater than $N$.\footnote{See, for instance, \cite{CS} for some typical numbers.} \\

\subsection{The G-degree of an invariant}\label{GDI}
A G-degree is a positive integer associated to a connected tensor invariant\footnote{As said before, the G-degree is also associated with Feynman bipartite graphs.}. It is a key concept in tensor models since the $1/N$ expansion of the partition function is driven by it. The leading order corresponds to invariants with degree 0 (melons) and subleading orders appear with an increasing value of their G-degree. Its definition and construction can be found in the original paper \cite{1N3} or in the more recent paper \cite{Casali:2017tfh}, where interesting features of the G-degree are analysed. 

Given a $d$-invariant $\gamma$ (a connected $d$-graph) of order $n$, the G-degree $\omega(\gamma)$ is computed as
\be \label{gdegree1}
\omega(\gamma)=\frac{(d-2)!}{2}\Big(d-1+\frac{(d-1)(d-2)}{2}n-\sum_{r,s}g_{rs}(\gamma)\Big),
\ee
where $g_{rs}(\gamma)$ is calculated as follows. First, given the graph $\gamma$ delete all the color lines except for the colors $r$ and $s$. Then count the number of connected components that these two color lines form. The sum in  \eqref{gdegree1} runs over all couples of colors.

The G-degree is very easy to compute when one takes the permutations notation for invariants $\gamma=\mathcal{O}_{\vec{\alpha}}$. The number of connected components of the reduced graph  to a given couple of color lines $(rs)$ is simply the number of cycles of the permutation $\alpha_r\alpha_s^{-1}$, which we call $C(\alpha_r\alpha_s^{-1})$. So, the G-degree is
\be \label{gdegree2}
\omega(\mathcal{O}_{\vec{\alpha}})=\frac{(d-2)!}{2}\Big(d-1+\frac{(d-1)(d-2)}{2}n-\sum_{i<j} C(\alpha_i\alpha_j^{-1})\Big).
\ee
Notice that the functions $C(\alpha_i\alpha_j^{-1})$ are invariant under the change $\alpha_k\to \sigma\alpha_k\tau$ for all $k$, with $\sigma,\tau\in S_n$, 
\be
C(\sigma\alpha_i\tau\tau^{-1}\alpha_j^{-1}\sigma^{-1})=C(\sigma\alpha_i\alpha_j^{-1}\sigma^{-1})=C(\alpha_i\alpha_j^{-1}).
\ee
This is a consistency check, since we know that $\mathcal{O}_{\vec{\alpha}}\sim \mathcal{O}_{\sigma\cdot\vec{\alpha}\cdot\tau}$, with $\sigma$ and $\tau$ acting diagonally. So, the G-degree should also be invariant under the change. As far as I know, equation \eqref{gdegree2} has not been found before and seems very useful for calculations. 

Let us see a couple of examples. Consider the observable $\mathcal{O}_{\alpha,1,\dots,1}$, where $\alpha$ is a 1-cycle permutation. It is easy to see that it is a melon. The number of cycles of the identity is $n$. Using simple combinatorics we get
\be
\omega(\mathcal{O}_{\alpha,1,\dots,1})=\frac{(d-2)!}{2}\Big(d-1+\frac{(d-1)(d-2)}{2}n-\binom{d-1}{2}n-(d-1)\Big)=0.
\ee
As a second example, and for later use, let me compute the G-degree of $\mathcal{O}_{\alpha,\sigma,1,\dots,1}$, where $\alpha$ is again a 1-cycle and $\sigma$ is a transposition, which has $n-1$ cycles. Using the fact that $C(\alpha \sigma^{-1})=2$, we may write
\be
\sum_{i<j} C(\alpha_i\alpha_j^{-1})=\binom{d-2}{2}n+(d-2)(n-1)+(d-2)+2,
\ee
and so 
\be
\omega(\mathcal{O}_{\alpha,\sigma,1,\dots,1})=\frac{(d-2)!}{2}(d-3).
\ee
Thus, $\mathcal{O}_{\alpha,\sigma,1,\dots,1}$ is melonic only for $d=3$.

\subsection{Expectation values in terms of G-degrees}\label{EVGD}
The expectation value of an observable \eqref{expectation} can be expressed in terms of the G-degree of graphs. There are $n!$ Wick contractions in the sum. Let us define the $(d+1)$-graph 
\be\label{sigmagraph}
\mathcal{O}_{\vec{\alpha},\sigma}=\mathcal{O}_{\alpha_1,\dots,\alpha_d,\sigma}.
\ee
We can build $n!$ graphs like \eqref{sigmagraph}, and I will associate each of them with a Wick contraction.
It is clear that
\be
\frac{2}{(d-2)!}\omega(\mathcal{O}_{\vec{\alpha}})-\frac{2}{(d-1)!}\omega(\mathcal{O}_{\vec{\alpha},\sigma})=C(\sigma \alpha_1)+\dots +C(\sigma \alpha_d)-1-(d-1)n,
\ee
so,
\be\label{expectationgdegree}
\la \mathcal{O}_{\vec{\alpha}} \lb=N^{1+(d-1)(n-1)}\sum_{\sigma\in S_n}N^{\frac{2}{(d-2)!}\omega(\mathcal{O}_{\vec{\alpha}})-\frac{2}{(d-1)!}\omega(\mathcal{O}_{\vec{\alpha},\sigma})}.
\ee
It is up to the taste of the reader to use \eqref{expectation} or \eqref{expectationgdegree} in the computation of the expectation value of an observable. 
In equation \eqref{expectationgdegree} it is clear how $(d+1)$-graphs (or observables) $\mathcal{O}_{\vec{\alpha},\sigma}$ appear in the computation of the expectation value of any order-$d$ tensor observable. In the usual manner, each $(d+1)$-graph encodes a triangulation, via gluing faces, of a $d$-dimensional PL manifold. This is the precise way in which tensor observables relate to triangulations.

\section{Double limit}\label{DL}
Tensor models are expected to encode classical gravity at the limit $N\to \infty$. The number of tensors that the invariants are built on, $2n$, is also supposed to be large at the continuum limit, where Einstein gravity is expected to be recovered. These two limits are usually taken uncorrelated. The usual picture for the emergence of space from tensor models is the following. Tensor invariants can be put in contact with triangulations: the expectation value of every order-$d$ tensor invariant built as a composite of $2n$ tensors produces $(d+1)$-graphs (see eq. \eqref{expectationgdegree}), each one encoding a $d$-dimensional  PL manifold made of $2n$ simplices. Basically, the pattern of contractions among  indices of the same color in a given graph dictates how the faces of the simplices must be glued in order to form the PL manifold. The tensor invariant is the ``skeleton'' of the triangulation. Interestingly, it was proven in \cite{Bonzom:2011zz} that the statistics of the tensor model relates to that of Regge triangulations as long as we make the right identifications. In the Regge context, smooth manifolds are recovered as we increase the number of simplices $n$ and decrease the length $l$ of  the side of the simplex cell.  This is indeed the way the continuum limit is achieved when recovering Einstein gravity from Regge (discrete) calculus \cite{regge}. For tensors, the limiting procedure is somehow vague in this usual picture since
\begin{enumerate}
    \item No length appears in the tensor model to account for the $l\to 0$ limit. So far, the tensor model seems to provide only the triangulations of the Regge theory.
    \item There is not a clear role for the parameter $N$, the size of the tensors, apart from the fact that it should be large to make contact with gravity.
\end{enumerate}

In this section I am going to discuss the role of $N$ and its connection with $l$ (and $n$) by identifying them with the bundle structure of the limiting manifold. Then,  the above caveats get smoothed out. 

First, let us remember how the limiting procedure of triangulations is achieved in the context of Regge calculus. The limiting manifold is obtained as a limit of a sequence of triangulations with more and more simplices involved. When proving that the Regge action reproduces the Einstein-Hilbert action in the continuum limit, the sequence of triangulations was taken backwards: one starts with the limiting manifold and ``inscribe'' a triangulation of length $l$ on it by marking a selection of points on the limiting manifold \cite{regge}. Triangulations can always be chosen equilateral for each $l$, this election is tantamount to partially fixing the gauge in gravity. This method of discretizing the space seems to be a convenient choice, and we will adopt it here. From now on, the PL manifold with $2n$ simplices (and also $2n$ vertices) will be interpreted as ``anchored'' on the limiting $d$-dimensional manifold at $2n$ points. 


If a certain Riemannian geometry with volume $V=\int \sqrt{g}$ is to be obtained at the limiting process it is clear that the volume of the PL manifold $V_{PL}$ must approach $V$ in the sense that, in the process of refinement, there should be $n_0$ such that $n>n_0\to |V-V_{PL}|<\delta$.\,This consideration tells us that for large $n$
\be\label{limit1}
n\, l^d\sim V,
\ee
 where $V$ is fixed since it is the volume of the limiting manifold, and $l$ is the length of the simplices. So, the limits $l\to 0$ and $n\to \infty$ should not be taken separately.

As said above, tensor models do not involve a natural length in their formulation but a parameter $N$ that measures the size of the tensor. Actually, when relating tensor invariants with triangulations no mention of the size of the tensor $N$ or their specific $N^d$ entries was made. The triangulation is only ruled by the contractions of the color indices of the $2n$ tensors which build the invariant. One might wonder what information is stored in the particular entries of the tensor and if they are relevant when making contact with gravity. In the usual picture, each of the $d$ indices of the tensor relates vaguely to a coordinate at the continuum limit. That way a model of tensors of order $d$ is expected to describe gravity in $d$ dimensions. The finite size $N$ of the tensors imply that the  coordinates are discretized.

In order to flesh out the previous idea we are taking a slight  departure from the usual picture above described. The crucial point is the number of degrees of freedom that the tensor has. As discussed in section \ref{R1}, the canonical decomposition \eqref{tensordecomp} is a natural way of organizing the degrees of freedom of a tensor. For instance, rank-1 tensors can be written as a tensor product of $d$ complex $N$-vectors, in which case the number of {\it real} degrees of freedom is simply $2dN$, up to rescaling.   The obvious fact is that each term of the sum in the canonical decomposition encodes the information of $d$ complex $N$-vectors. Alternatively the array can be rotated and each term of the sum may be seen as $N$ complex $d$-uplas, or  complex $d$-vectors placed in a grid with $N$ sites. This is the same as counting real $d$-vectors placed in a grid with $2N$ sites. The reason for organizing the degrees of freedom this way is because in $d$-dimensional gravity $d$-vectors (and no $N$-vectors) occur. Actually, in gravity, $d$-vectors live naturally at the tangent space of each point of the manifold. Thus, it seems reasonable to associate the information encoded in a rank-1 tensor $T$ to a (discretized) real vector field, each vector living at what will become  the tangent space of the manifold at a point, when the continuum limit is taken. Notice that in the new arrangement, each unitary group $U_k(N)$ of the gauge group $U(N)^{\times d}$ acts on the $k$-upla of all the $N$ $d$-vectors, preluding what will be reparametrizations at the continuum limit.    

As commented above the rank of the tensor is irrelevant for the purpose of encoding triangulations. We can perfectly use a rank-1 tensor to build the invariant that relates to a given triangulation. let us first consider rank-1 tensors. In this case we are left with a PL manifold anchored at $2n$ points of a $d$-dimensional limiting manifold on the one hand, and a collection of $2N$ real $d$-vectors encoded in the rank-1 tensor $T$ on the other. In the continuum limit each point has its own tangent space, a fact that must persist in any discretization. Thus,
\be\label{nequalN}
N\sim n.
\ee
Since $d$-vectors live in the tangent space, the statement \eqref{nequalN} is equivalent to saying that every point in a manifold has its own tangent space. Now, with ``double limit'' in this context I mean a limit where $n,N\to \infty$ but keeping  \eqref{nequalN}.

In summary, the degrees of freedom of a rank-1 tensor are interpreted as a  discretized $d$-vector field at the anchor points. Generic tensors of rank $R(T)$ are then interpreted as a collection of such configurations on the grid with $2n$ sites, where the PL manifold is anchored. Remember that the number $R(T)$ is usually greater than $N$. So, I expect that for large $N$ the tensor degrees of freedom  of a tensor of generic rank actually build a fiber at each site of the grid, leading to a vector bundle at the continuum limit. 

Now, the relation \eqref{limit1} turns into a relation between the unit length $l$ in the Regge context and the size of the tensor $N$ as
\be
N\sim\frac{V}{l^d},
\ee
which must hold as  the continuum limit is approached.\\

The discussion of this section is mainly based on organizing and counting degrees of freedom in tensors. In this sense, the analysis is still preliminary. However, these considerations have allowed me to argue that the tensor degrees of freedom encode vector fields in the tangent space of the manifold. Following this logic, the main result, expressed in \eqref{nequalN}, is the statement that each point of the manifold carries its own tangent space, which is the main message of the argument.

\section{The defeat of melons}\label{DoM}
 The relation \eqref{nequalN} in the double limit $n,N\to \infty$ has important implications. One of the main consequences is the failure of melonic diagrams to dominate the $1/N$ expansion of the theory. 
 A key concept for stating the dominance of melon graphs at large $N$ is the Gurau degree or G-degree.

 From \eqref{expectationgdegree} it is clear that melon graphs, that is,  the graphs for which $\omega(\mathcal{O}_{\vec{\alpha},\sigma})=0$ dominate when computing the expectation value of an observable in the large $N$ limit. Especially, since  $\omega(\mathcal{O}_{\vec{\alpha},\sigma})=0\longrightarrow \omega(\mathcal{O}_{\vec{\alpha}})=0$. In other words, no melon $d+1$-graph appears in the computation of the expectation value of a non-melonic observable \cite{Bonzom:2012hw}.  So, melonic terms in the two-point function are the relevant ones at large $N$.

As anticipated in the title of this section, this argument breaks down when we consider the double limit \eqref{nequalN}. Basically what happens is that although the individual non-melonic terms are subleading at large $N$, their number grow so much as to overcome the total sum.   \\
Let us consider the invariant $\mathcal{O}_{\alpha,1,\dots,1}$, where $\alpha$ is a 1-cycle permutation. As discussed before, this is a melonic observable, so $\omega(\mathcal{O}_{\alpha,1,\dots,1})=0$. According to \eqref{expectationgdegree}
\be
\la \mathcal{O}_{\alpha,1,\dots,1} \lb=N^{1+(d-1)(n-1)}\sum_{\sigma\in S_n}N^{-\frac{2}{(d-1)!}\omega(\mathcal{O}_{\alpha,\sigma,1,\dots,1})}.
\ee
Note that for $\mathcal{O}_{\alpha,\sigma,1,\dots,1}$ to be melonic $\sigma$ must be 1. Thus there is only one melonic term in the sum and its contribution is
\be
\la \mathcal{O}_{\alpha,1,\dots,1} \lb_{\text{melonic}}=N^{1+(d-1)(n-1)}.
\ee
This is the leading order for observables for which $n\ll N$.

Let us compute the contribution of the terms where $\sigma$ is a transposition. There are $\binom{n}{2}$ terms of this type, and their G-degree is $\frac{(d-1)!}{2}(d-2)$. Taking the double limit \eqref{nequalN}, the ratio between melonic and non-melonic contribution of the transpositions is
\be
\frac{\la \mathcal{O}_{\alpha,1,\dots,1}\lb_{\text{trans}}}{\la \mathcal{O}_{\alpha,1,\dots,1} \lb_{\text{melonic}}}=\sum_{\sigma=(ij)}N^{-(d-2)}\sim N^{-(d-2)}\binom{n}{2}\sim  N^{4-d} ,
\ee
where I have used $n\sim N$. This indicates that for order 4, for instance, the contribution of the non-melonic transposition diagrams is of the same order as the melonic contribution. For order 3 the non-melonic transpositions diagrams actually overcome the total sum. In both cases, the non-melonic contribution cannot be neglected any longer. 

This is an easy counter-example of the melonic dominance, it will be more elaborated in the future, but the conclusion is that, at the double limit \eqref{nequalN}, melonic contributions to the expectation value of an observable are not necessarily the leading order. So, techniques to sum {\it all} contributions must be develop in the future. 

Analogous arguments to these were used in the context of holography and matrix theories, when studying dual operators to giant gravitons, for instance. In that scenario the observables needed to be large and grew with $N$ as well, and that made non-planar diagrams dominate \cite{Balasubramanian:2001nh}. Techniques to sum all contributions needed to be developed \cite{Corley:2001zk,Bhattacharyya:2008rb,Bhattacharyya:2008xy,Kimura:2007wy,Brown:2007xh,deMelloKoch:2011vn,Koch:2012sf}.  


\section{Conclusion and outlook}
In this paper I have studied the limits $N,n\to \infty$, where tensor models are expected to describe gravity. This link is made by the connection that tensor models have with Regge calculus, and the fact that the continuum limit of the Regge theory leads to Einstein gravity. The physical meaning of $n$ in this picture is well-established: an invariant built on $2n$ tensors corresponds to a PL manifold with $2n$ simplices. The specific triangulation is encoded in the pattern of contractions that define the invariant. 

The physical meaning of $N$ and the degrees of freedom of the tensor are not so clear in the literature and one of the motivations of this paper is to cover this issue. By arranging the degrees of freedom of tensors in a convenient way, the canonical decomposition \eqref{tensordecomp}, they seem to encode discretized vector fields. At the continuum limit these vectors live in the tangent space of the manifold.  Thus, I claim that the tensor degrees of freedom build a vector bundle at the continuum limit, whose sections could be thought of as coordinate systems. Based on this picture and the fact that every point in a manifold has its own tangent space I argue that $N\sim n$ if the tensor model is to be used to describe gravity. 

The fact that the limits $N,n\to \infty$ are correlated have implications in the computation of the expectation values of the observables of the theory. Because the number of diagrams grows rapidly with $n$, melon diagram contributions, which are the leading order in $N$, can get overcome by other subleading but more numerous diagrams. I give a couple of examples where this actually happens.   So far, it seems that there is more than melons in the fruit shop.

The analysis performed in this paper is based on counting and arranging degrees of freedom. In this sense it is still preliminary. I hope to flesh out these ideas and find proofs of the main claims of this paper in future works.

\vspace{0.2cm}
\noindent
{\bf Acknowledgment}

\noindent
I would like to thank Robert de Mello Koch and Antonio Segu\'i for their valuable comments. I am also grateful to the OIST Mini Symposium ``Holographic Tensors'' and its interesting talks, where I found true inspiration and some of my dreams on these topics became ideas. This work is supported by Universidad de Zaragoza.

\end{document}